\documentstyle[amstex,amssymb]{article}

\catcode`\@=11
\def\numberbysection{\@addtoreset{equation}{section}
        \def\theequation{\thesection.\arabic{equation}}}
\numberbysection

\begin{document}

\newlength{\lno} \lno1.75cm \newlength{\len} \len=\textwidth%
\addtolength{\len}{-\lno}

\setcounter{page}{0}

\baselineskip8mm \renewcommand{\thefootnote}{\fnsymbol{footnote}} \newpage %
\setcounter{page}{0}

\begin{titlepage}     
\vspace{0.5cm}
\begin{center}
{\Large\bf Temperley-Lieb K- matrices}\\
\vspace{1cm}
{\large \bf  A. Lima-Santos } \\
\vspace{1cm}
{\large \em Universidade Federal de S\~ao Carlos, Departamento de F\'{\i}sica \\
Caixa Postal 676, CEP 13569-905~~S\~ao Carlos, Brasil}\\
\end{center}
\vspace{1.2cm}

\begin{abstract}
This work concerns to the studies of  boundary integrability of the vertex models  from representations of the Temperley-Lieb 
algebra associated with the quantum group  ${\cal U}_{q}[X_{n}]$ for the affine Lie algebras
 $X_{n}$ = $A_{1}^{(1)}$, $B_{n}^{(1)}$, $C_{n}^{(1)}$ and $D_{n}^{(1)}$. 

A systematic computation method is used to constructed  solutions of the boundary Yang-Baxter equations.  
We find a  $2n^{2}+1$  free parameter solution  for $A_{1}^{(1)} $ spin-$(n-1/2)$ and $ C_{n}^{(1)}$ 
vertex models.  It turns that for  $A_{1}^{(1)} $ spin-$n$, $ B_{n}^{(1)}$  and $D_{n}^{(1)}$ vertex models, the solution 
has $2n^{2}+2n+1$  free parameters.
\end{abstract}
\vfill
\begin{center}
\small{\today}
\end{center}
\end{titlepage}

\baselineskip7mm

\newpage

\section{Introduction}

The search for integrable models through solutions of the Yang--Baxter
equation \cite{Baxter, KIB, AAR}%
\begin{equation}
{\cal R}_{12}(u-v){\cal R}_{13}(u){\cal R}_{23}(v)={\cal R}_{23}(v){\cal R}%
_{13}(u){\cal R}_{12}(u-v)  \label{int.1}
\end{equation}%
has been performed by the quantum group approach in \cite{KR}. In this way,
the ${\cal R}$-matrices corresponding to vector representations of all
nonexceptional affine Lie algebras have been determined in \cite{Jimbo}.

A similar approach is desirable for finding solutions of the boundary
Yang--Baxter equation \cite{Chere, Skly} where the boundary weights follow
from $K$-matrices which satisfy a pair of equations, namely the reflection
equation%
\begin{equation}
{\cal R}_{12}(u-v)K_{1}^{-}(u){\cal R}%
_{12}^{t_{1}t_{2}}(u+v)K_{2}^{-}(v)=K_{2}^{-}(v){\cal R}%
_{12}(u+v)K_{1}^{-}(u){\cal R}_{12}^{t_{1}t_{2}}(u-v)  \label{int.2}
\end{equation}%
and the dual reflection equation%
\begin{eqnarray}
&&{\cal R}_{12}(-u+v)\left( K_{1}^{+}\right) ^{t_{1}}(u)M_{1}^{-1}{\cal R}%
_{12}^{t_{1}t_{2}}(-u-v-2\rho )M_{1}\left( K_{2}^{+}\right) ^{t_{2}}(v)= 
\nonumber \\
&&\left( K_{2}^{+}\right) ^{t_{2}}(v)M_{1}{\cal R}_{12}(-u-v-2\rho
)M_{1}^{-1}\left( K_{1}^{+}\right) ^{t_{1}}(u){\cal R}%
_{12}^{t_{1}t_{2}}(-u+v).  \label{int.3}
\end{eqnarray}%
In this case duality supplies a relation between $K^{-}$ and $K^{+}$ \cite%
{MN1}%
\begin{equation}
K^{+}(u)=K^{-}(-u-\rho )^{t}M,\qquad M=V^{t}V  \label{int.4}
\end{equation}%
Here $t$ denotes transposition and $t_{i}$ denotes transposition in the i-th
space. $V$ is the crossing matrix and $\rho $ the crossing parameter, both
being specific to each model \cite{Bazha}.

With this goal in mind, the study of boundary quantum groups was initiated
in \cite{MN2}. This study have been used to determine $A_{1}^{(1)}$
reflection matrices for arbitrary spin \cite{DN}, and the $A_{2}^{(2)}$ and
some $A_{n}^{(1)}$ reflection matrices were derived again in \cite{Nepo}.
Reflection solutions from ${\cal R}$-matrices corresponding to vector
representations of Yangians and super-Yangians were presented in \cite{Arnau}%
. However, as observed in \cite{Nepo}, an independent systematic method of
constructing the boundary quantum group generators is not yet available. In
contrast to the bulk case \cite{Jimbo}, one cannot exploit boundary affine
Toda field theory, since appropriate classical integrable boundary
conditions are not yet known \cite{BCDR}.

However, the algebraic structures related to reflection equations are
well-known \cite{KSkly} and a boundary quantum group approach was recently
used in \cite{Avan} to derive the classification of the constant $K$-matrix
(without spectral parameter) solutions for the Temperley-Lieb ({\small TL})
models. The main result, already pointed in \cite{Kulish}, is that for a
given constant {\small TL} ${\cal R}$-matrix, the corresponding constant $K$%
-matrix satisfies a quadratic relation:%
\begin{equation}
qK^{2}+c_{1}K+(q+q^{-1})^{-1}(c_{1}^{2}+qc_{2})I=0  \label{int.5}
\end{equation}%
with appropriate central elements $c_{1}$ and $c_{2}$.

From this result, the Yang-Baxterization procedure, as used in \cite{[11],
[12], [13]}, allows to obtain spectral-parameter dependent reflection
matrices:%
\begin{equation}
K(u)=u^{2}K-\frac{1}{u^{2}}K^{-1}+cI  \label{int.6}
\end{equation}%
with an arbitrary central element $c$.

Independently, there has been an increasing amount of effort towards the
understanding of two-dimensional integrable theories with boundaries via
solutions of the functional equation (\ref{int.2}). In field theory,
attention is focused on the boundary $S$ matrix \cite{GZ, FK}. In
statistical mechanics, the emphasis has been laid on deriving all solutions
of (\ref{int.2}) because different $K$-matrices lead to different
universality classes of surface critical behavior \cite{Batch1} and allow
the calculation of various surface critical phenomena, both at and away from
criticality \cite{Batch2}.

Although being a hard task, the direct computation has been used to solve (%
\ref{int.2}). For instance, we mention the solutions with ${\cal R}$ matrix
based in non-exceptional Lie algebras \cite{DeVega, RLS} and superalgebras 
\cite{Doikou1, LS}. The regular $K$-matrices for the exceptional ${\cal U}%
_{q}[G_{2}]$ vertex model were obtained in \cite{LSM}$.$ Many diagonal
solutions for face and vertex models associated with affine Lie algebras
were presented in \cite{Batch2}. For {\small A--D--E} interaction-round face
({\small IRF}) models, diagonal and some non-diagonal solutions were
presented in \cite{BP}. Reflection matrices for Andrews--Baxter--Forrester
models in the {\small RSOS/SOS} representation were presented in \cite{Ahn}.
Apart from these $c$-number solutions of the reflection equations there must
also exist non trivial solutions that include boundary degree of freedom as
were derived for the sine-Gordon theory in \cite{BK} and the projected $K$%
-matrices \cite{FS}.

Motived by the results presented in \cite{LS3} we will again touch this
issue in order to include once more the {\small TL} lattice models \cite{TL}
arising from the quantum group ${\cal U}_{q}[X_{n}]$ for $%
X_{n}=A_{1}^{(1)},B_{n}^{(1)},C_{n}^{(1)}$ and $D_{n}^{(1)}$\cite{Batch3}.

The {\small TL} algebra is very useful in the study of two dimensional
lattice statistical mechanics. It provided an algebraic framework for
constructing and analyzing different types of integrable lattice models,
such as $Q$-state Potts model, {\small IRF} model, $O(n)$ loop model,
six-vertex model, etc. \cite{Martin1}.

We have organized this paper as follows. In Section $2$ the model is
presented, in Section $3$ we choose the reflection equations and their
solutions. The Section $4$ is reserved for the conclusion.

\section{The model}

From the representation of the {\small TL} algebra, one can build solvable
vertex models with the $R$ operator defined by 
\begin{equation}
R(u)=x_{1}(u)I+x_{2}(u){\cal U},  \label{mod.1}
\end{equation}%
where $I$ is the identity operator and ${\cal U}$ is the {\small TL}
projector. Here $u$ is the spectral parameter and the anisotropic parameter $%
\eta $ is choose so that%
\begin{eqnarray}
x_{1}(u) &=&\frac{\sinh (\eta -u)}{\sinh \eta },\quad x_{2}(u)=\frac{\sinh u%
}{\sinh \eta },\quad  \nonumber \\
2\cosh \eta &=&{\rm Tr}\text{ }{\cal U}\text{.}  \label{mod.2}
\end{eqnarray}%
Setting 
\begin{equation}
R_{j}(u)=1\otimes \cdots 1\otimes \underset{j,j+1}{\underbrace{R(u)}}\otimes
1\cdots \otimes 1  \label{mod;3}
\end{equation}%
one can show that the Yang-Baxter equation 
\begin{equation}
R_{j+1}(u)R_{j}(u+v)R_{j+1}(v)=R_{j}(v)R_{j+1}(u+v)R_{j}(u)  \label{mod.4}
\end{equation}%
is valid due to the definition relations of the {\small TL} algebra%
\begin{eqnarray}
{\cal U}_{j}^{2} &=&2\cosh \eta \ {\cal U}_{j}  \nonumber \\
{\cal U}_{j}{\cal U}_{j\pm }{\cal U}_{j} &=&{\cal U}_{j}  \nonumber \\
{\cal U}_{i}{\cal U}_{j} &=&{\cal U}_{j}{\cal U}_{i}\qquad |i-j|>1
\label{mod.5}
\end{eqnarray}

For the affine Lie algebras $A_{1}^{(1)}$, $B_{n}^{(1)}$, $C_{n}^{(1)}$ and $%
D_{n}^{(1)}$ {\it i.e., }the $q$-deformations of the spin-$s$ representation
of $sl(2)$ and the vector representation of $so(2n+1)$, $sp(2n)$ and $so(2n)$%
, the corresponding {\small TL} projector, using the notation and results of 
\cite{Batch1}, has the form 
\begin{equation}
{\cal U}=\sum_{i,j=1}^{N}\varepsilon (i)\varepsilon (j)\ q^{-<{\bf \epsilon }%
_{i}+{\bf \epsilon }_{j},\overset{\_}{\rho }>}{\rm e}_{i,j}\otimes {\rm e}%
_{i^{\prime },j^{\prime }}  \label{mod.6}
\end{equation}%
where ${\rm e}_{i,j}$ is the matrix unit (${\rm e}_{i,j}v_{k}=\delta
_{j,k}v_{i}$) and we have used the conjugated index $a^{\prime }=N+1-a$.

Here, one has to take into account the set of orthonormal vectors $<{\bf %
\epsilon }_{i},{\bf \epsilon }_{j}>=\delta _{i,j}$ , the sign $\varepsilon
(i)$ and $\overset{\_}{\rho }$, the half-sum of positive roots of the $q$%
-deformed affine Lie algebras in order to write explicitly the {\small TL}
projector for each model:

\begin{itemize}
\item $A_{1}^{(1)}$: The ${\cal U}_{q}[sl(2)]$ spin-$s$ Temperley-Lieb model%
\[
{\cal U}=\sum_{i=1}^{N}\sum_{j=1}^{N}(-1)^{i+j}q^{i+j-N-1}{\rm e}%
_{i,j}\otimes {\rm e}_{i^{\prime },j^{\prime }} 
\]%
\begin{equation}
2\cosh \eta =[2s+1]  \label{mod.7}
\end{equation}%
where $N=2s+1$ $(s=\frac{1}{2},1,\frac{3}{2},2,...)$. We remark the use of
the quantum number notation $[n]=(q^{n}-q^{-n})/(q-q^{-1})$ in the trace of
the ${\cal U}$ projectors.

\item $B_{n}^{(1)}(n\geq 2)$ : The ${\cal U}_{q}[so(2n+1)]$ Temperley-Lieb
model%
\begin{eqnarray*}
{\cal U} &=&\sum_{i=1}^{n}\sum_{j=1}^{n}q^{i+j-2n-1}{\rm e}_{i,j}\otimes 
{\rm e}_{i^{\prime },j^{\prime }}-\sum_{i=1}^{n}q^{i-n-1/2}{\rm e}%
_{i,n+1}\otimes {\rm e}_{i^{\prime },n+1} \\
&&+\sum_{i=1}^{n}\sum_{j=n+2}^{2n+1}q^{i+j-2n-2}{\rm e}_{i,j}\otimes {\rm e}%
_{i^{\prime },j^{\prime }}-\sum_{j=1}^{n}q^{j-n-1/2}{\rm e}_{n+1,j}\otimes 
{\rm e}_{n+1,j^{\prime }} \\
&&+{\rm e}_{n+1,n+1}\otimes {\rm e}_{n+1,n+1}-\sum_{j=n+2}^{2n+1}q^{j-n-3/2}%
{\rm e}_{n+1,j}\otimes {\rm e}_{n+1,j^{\prime }} \\
&&+\sum_{i=n+2}^{2n+1}\sum_{j=1}^{n}q^{i+j-2n-2}{\rm e}_{i,j}\otimes {\rm e}%
_{i^{\prime },j^{\prime }}-\sum_{i=n+2}^{2n+1}q^{i-n-3/2}{\rm e}%
_{i,n+1}\otimes {\rm e}_{i^{\prime },n+1} \\
&&+\sum_{i=n+2}^{2n+1}\sum_{j=n+2}^{2n+1}q^{i+j-2n-3}{\rm e}_{i,j}\otimes 
{\rm e}_{i^{\prime },j^{\prime }}
\end{eqnarray*}%
\begin{equation}
2\cosh \eta =\frac{[2n-1][n+\frac{1}{2}]}{[n-\frac{1}{2}]}  \label{mod.8}
\end{equation}

\item $C_{n}^{(1)}(n\geq 1)$: The ${\cal U}_{q}[sp(2n)]$ Temperley-Lieb model%
\begin{eqnarray*}
{\cal U} &=&\sum_{i=1,}^{n}\sum_{j=1}^{n}q^{i+j-2n-2}{\rm e}_{i,j}\otimes 
{\rm e}_{i^{\prime },j^{\prime
}}-\sum_{i=1,}^{n}\sum_{j=n+1}^{2n}q^{i+j-2n-1}{\rm e}_{i,j}\otimes {\rm e}%
_{i^{\prime },j^{\prime }} \\
&&-\sum_{i=n+1,}^{N}\sum_{j=1}^{n}q^{i+j-2n-1}{\rm e}_{i,j}\otimes {\rm e}%
_{i^{\prime },j^{\prime }}+\sum_{i=n+1,}^{2n}\sum_{j=n+1}^{2n}q^{i+j-2n}{\rm %
e}_{i,j}\otimes {\rm e}_{i^{\prime },j^{\prime }}
\end{eqnarray*}%
\begin{equation}
2\cosh \eta =\frac{[n][2n+2]}{[n+1]}  \label{mod.9}
\end{equation}

\item $D_{n}^{(1)}(n\geq 3)$: The ${\cal U}_{q}[so(2n)]$ Temperley-Lieb model%
\begin{eqnarray*}
{\cal U} &=&\sum_{i=1,}^{n}\sum_{j=1}^{n}q^{i+j-2n}{\rm e}_{i,j}\otimes {\rm %
e}_{i^{\prime },j^{\prime }}+\sum_{i=1}^{n}\sum_{j=n+1}^{2n}q^{i+j-2n-1}{\rm %
e}_{i,j}\otimes {\rm e}_{i^{\prime },j^{\prime }} \\
&&+\sum_{i=n+1}^{2n}\sum_{j=1}^{n}q^{i+j-2n-1}{\rm e}_{i,j}\otimes {\rm e}%
_{i^{\prime },j^{\prime }}+\sum_{i=n+1}^{2n}\sum_{j=n+1}^{2n}q^{i+j-2n-2}%
{\rm e}_{i,j}\otimes {\rm e}_{i^{\prime },j^{\prime }}
\end{eqnarray*}%
\begin{equation}
2\cosh \eta =\frac{[n][2n-2]}{[n-1]}  \label{mod.10}
\end{equation}
\end{itemize}

We also have to consider the permuted operator ${\cal R}=PR$ which is
regular satisfying {\small PT}-symmetry, unitarity and crossing symmetry%
\begin{eqnarray}
{\cal R}_{12}(0) &=&P,  \nonumber \\
{\cal R}_{12}^{t_{1}t_{2}}(u) &=&P{\cal R}_{12}(u)P={\cal R}_{21}(u), 
\nonumber \\
{\cal R}_{12}(u){\cal R}_{21}(-u) &=&x_{1}(u)x_{1}(-u)I,  \nonumber \\
{\cal R}_{21}(u) &=&\kappa (V\otimes 1){\cal R}_{12}^{t_{2}}(-u-\rho
)(V\otimes 1)^{-1}  \label{mod.12}
\end{eqnarray}%
where $\rho =-\eta $ is the crossing parameter, $\kappa =(-1)^{2s}$ for $%
A_{1}^{(1)}$, $\kappa =-1$ for $C_{n}^{(1)}$ and $\kappa =1$ for $%
B_{n}^{(1)} $ and $D_{n}^{(1)}$. $P$ is the permutation operator: $%
P(a\otimes b)=b\otimes a$ for any vectors $a,b$.

The crossing matrices $V$ for the {\small TL} models are specified in \cite%
{Batch1} by 
\begin{equation}
V_{i,j}=\varepsilon (i)\ q^{-<{\bf \epsilon }_{i},\rho >}\delta _{i^{\prime
},j}  \label{mod.13}
\end{equation}%
However, for the isomorphism (\ref{int.4}) we only need of the diagonal
matrix $M=V^{t}V$ for each model: 
\begin{equation}
M_{i,i}=q^{2i-2s-2}\qquad i=1,...,2s+1\quad {\rm for}\quad A_{1}^{(1)}\quad 
{\rm spin}-s  \label{M.1}
\end{equation}%
\begin{equation}
M_{i,i}=\left\{ 
\begin{array}{c}
q^{2i-2n-1}\quad 
\hfill%
i=1,...,n \\ 
1\quad 
\hfill%
i=n+1 \\ 
q^{2i-2n-3}\quad 
\hfill%
i=n+2,...,2n+1%
\end{array}%
\right. \quad {\rm for}\quad B_{n}^{(1)}  \label{M.2}
\end{equation}%
\begin{equation}
M_{i,i}=\left\{ 
\begin{array}{c}
q^{2i-2n-2}\quad 
\hfill%
i=1,...,n \\ 
q^{2i-2n}\quad 
\hfill%
i=n+1,...,2n%
\end{array}%
\right. \quad {\rm for}\quad C_{n}^{(1)}  \label{M.3}
\end{equation}%
and%
\begin{equation}
M_{i,i}=\left\{ 
\begin{array}{c}
q^{2i-2n}\quad 
\hfill%
i=1,...,n \\ 
1\quad 
\hfill%
i=n+1,n+2 \\ 
q^{2i-2n-2}\quad 
\hfill%
i=n+3,...,2n+1%
\end{array}%
\right. \quad {\rm for}\quad D_{n}^{(1)}  \label{M.4}
\end{equation}

The Hamiltonian limit 
\begin{equation}
R(u)={\cal I}+u(\alpha ^{-1}{\cal H}+\beta {\cal I})  \label{mod.14}
\end{equation}%
with $\alpha =\sinh \eta $ , $\beta =-\coth \eta $ leads to the quantum spin
chains 
\begin{equation}
{\cal H}=\sum_{k=1}^{N-1}{\cal U}_{k,k+1}+{\rm bt}  \label{mod.15}
\end{equation}%
where, instead of periodic boundary condition, we are taking into account
the existence of integrable boundary terms ${\rm bt}$ \cite{Skly}, derived
from the $K^{-}$ and $K^{+}$ matrices presented in the next sections.

\section{The reflection matrices}

In the reflection equation (\ref{int.2}) we remark the notation $%
K_{1}^{-}=K^{-}\otimes I$, $K_{2}^{-}=I\otimes K^{-}$, ${\cal R}_{12}={\cal R%
}$ and ${\cal R}_{12}^{t_{1}t_{2}}=P{\cal R}P$. For a given ${\cal R}$%
-matrix the unknown is the $N$ by $N$ matrix $K^{-}(u)$ satisfying the
normal condition $K^{-}(0)=I$. \ The dimension $N$ is equal to $2s+1,2n+1,2n$
and $2n+1$ for $A_{1}^{(1)},B_{n}^{(1)},C_{n}^{(1)}$ and $D_{n}^{(1)}$,
respectively.

Substituting%
\begin{equation}
K^{-}(u)=\sum_{i,j=1}^{N}k_{i,j}(u){\rm e}_{i,j}  \label{lkm.1}
\end{equation}%
and ${\cal R}(u)=P[x_{1}(u)I+x_{2}(u){\cal U}]$ into (\ref{int.2}), we will
have $N^{4}$ functional equations for the $k_{i,j}$ elements, many of them
not independent equations. In order to solve these functional equations, we
shall proceed as follows. First we consider the $(i,j)$ component of the
matrix equation (\ref{int.2}). By differentiating it with respect to $v$ and
taking $v=0$, we get algebraic equations involving the single variable $u$
and $N^{2}$ parameters%
\begin{equation}
\beta _{i,j}=\frac{dk_{i,j}(v)}{dv}|_{v=0}\text{ },\qquad i,j=1,2,...,N
\label{lkm.3}
\end{equation}

Analyzing the refection equations one can see that they possess a special
structure. Several equations exist involving only two non-diagonal elements.
They can be solved by the relations

\begin{equation}
k_{i,j}(u)=\frac{\beta _{i,j}}{\beta _{1,N}}k_{1,N}(u)\qquad (i\neq
j=\{1,2,...,N\})  \label{lkm.4}
\end{equation}%
We thus left with several equations involving two diagonal elements and $%
k_{1,N}(u)$. Such equations are solved by the relations%
\begin{equation}
k_{i,i}=k_{1,1}(u)+\left( \beta _{i,i}-\beta _{1,1}\right) \frac{k_{1,N}(u)}{%
\beta _{1,N}}\qquad (i=2,3,...,N).  \label{lkm.5}
\end{equation}%
Finally, we can use the equation $(1,N)$ in order to find the element $%
k_{1,1}(u)$:%
\begin{eqnarray}
k_{1,1}(u) &=&\frac{k_{1,N}(u)}{\beta _{1,N}[x_{2}(u)\cosh \eta +x_{1}(u)]}%
\left\{ \frac{x_{1}(u)x_{2}^{\prime }(u)-x_{1}^{\prime }(u)x_{2}(u)}{x_{2}(u)%
}\right.  \nonumber \\
&&\left. -\frac{1}{2}x_{1}(u)(\beta _{N,N}-\beta _{1,1}+\Psi _{1,N})-\frac{1%
}{2}x_{2}(u)\sum_{j=2}^{N}(\beta _{j,j}-\beta _{1,1})M_{j,j}\right\}
\label{k11}
\end{eqnarray}%
where $M_{j,j}$ are given in (\ref{M.1}--\ref{M.4}) and $\Psi _{1,N}$
belongs to the set of new relations of the parameters $\beta _{i,j}$ defined
by 
\begin{equation}
\Psi _{i,j}=\frac{1}{\beta _{i,j}}\sum_{k=2}^{N-1}\beta _{i,k}\beta
_{k,j}\qquad i\neq j=1,...,N  \label{nrel.1}
\end{equation}%
The prime in the Boltzmann weights $x_{i}(u)$ means its first derivative
with respect to $u$.

Now, substituting these expressions into the remained equations $(i,j)$, we
are left with factored equations of the form:%
\begin{equation}
F_{a}(\beta _{i,j})q^{p}x_{1}(u)x_{2}(u)k_{1,N}(u)=0  \label{ceqs}
\end{equation}%
where each factor $F_{a}(\beta _{i,j})$ don't depend on the weights $%
x_{i}(u) $ nor of the corresponding quantum group $q$-parameter. It means
that they are the same in all four models by we considering. Therefore all
computation used in \cite{LS3} give us a general procedure:

First, we collect all matrix element $(i,j)$ of (\ref{int.2}) in blocks of
four equations \cite{ALS}%
\begin{eqnarray}
B[i,j] &=&\left\{ (i,j),(j,i),(i^{\prime \prime },j^{\prime \prime
}),(j^{\prime \prime },i^{\prime \prime }))\right\}  \nonumber \\
i &=&1,...,N,\qquad j=i,...,i^{\prime \prime }  \label{lkm.6}
\end{eqnarray}%
where $a^{\prime \prime }=N^{2}+1-a$.

From the first equation of the blocks $B[j,N-1],$ $j=2,...,N-2$ one can fix $%
N-1$ diagonal parameters 
\begin{equation}
\beta _{j,j}=\beta _{1,1}+\Psi _{1,N}-\Psi _{j,N}\qquad j=2,3,...,N-1
\label{diag.1}
\end{equation}%
and the first equation of the block $B[N,N+1]$ fixes the parameter $\beta
_{N,N}$\ 
\begin{equation}
\beta _{N,N}=\beta _{1,1}+\Psi _{1,N-1}-\Psi _{N-1,N}  \label{diag.2}
\end{equation}%
All equations from the block $B[1,k]$ to the block $B[N-1,k]$ are now
substituted by $N(N-1)/2$ symmetric relations%
\begin{equation}
\Psi _{j,i}=\Psi _{i,j},\quad j>i  \label{sym}
\end{equation}%
and$\ 2(N-3)$ relations involving four $\Psi _{i,j}$ functions%
\begin{eqnarray}
\Psi _{2,j} &=&\Psi _{2,3}+\Psi _{1,j}-\Psi _{1,3},\quad \Psi _{3,j}=\Psi
_{2,3}+\Psi _{1,j}-\Psi _{1,2,}  \nonumber \\
j &=&4,...,N,  \label{psi.1}
\end{eqnarray}%
The remained equations contained in the block $B[N,k]$ are rewritten by $%
(N-3)(N-4)/2$ relations involving six $\Psi _{i,j}$ functions 
\begin{eqnarray}
\Psi _{i,j} &=&\Psi _{1,i}+\Psi _{1,j}+\Psi _{2,3}-\Psi _{1,2}-\Psi _{1,3}, 
\nonumber \\
\qquad i &=&4,...,N-1,\quad j=i+1,...,N  \label{psi.2}
\end{eqnarray}%
and $2N-3$ relations involving the diagonal $\beta _{k,k}$ parameters, $\Psi
_{1,N}$ and a new function $\Theta _{j,j}$, 
\begin{eqnarray}
\Theta _{j,j} &=&\Theta _{N,N}+(\beta _{N,N}-\beta _{j,j})(\beta
_{j,j}-\beta _{1,1}-\Psi _{1,N}),\quad j=2,3,...,N-1,  \nonumber \\
\Theta _{j^{\prime },j^{\prime }} &=&\Theta _{1,1}+(\beta _{1,1}-\beta
_{j^{\prime },j^{\prime }})(\beta _{j^{\prime },j^{\prime }}-\beta
_{N,N}-\Psi _{1,N}),\quad j=2,3,...,N-1,  \nonumber \\
\Theta _{N,N} &=&\Theta _{1,1}-(\beta _{1,1}-\beta _{N,N})\Psi _{1,N},
\label{theta.1}
\end{eqnarray}%
where $j^{\prime }=N+1-j$ and 
\begin{equation}
\Theta _{j,j}=\sum_{k\neq j}\beta _{j,k}\beta _{k,j}  \label{theta.2}
\end{equation}

From (\ref{sym}) to (\ref{theta.1}) one can account $N^{2}-3$ constraint
equations but, after the substitution of the relations (\ref{diag.1}) and (%
\ref{diag.2}) into (\ref{theta.1}), we only need to look at the symmetric
relations (\ref{sym}). However, for computational convenience, we added all
relations with four $\Psi _{i,j}$ functions (\ref{psi.1}). Therefore, our
final task is look for solutions of $N(N-2)$ constraint equations.

From these relations we have fixed $N(N-2)/2-1/2$ parameters $\beta $ for
the {\small TL} models with $N$ odd and $N(N-2)/2-1$ parameters for the 
{\small TL} models with $N$ even.

Taking into account that the parameter $\beta _{1,1}$ is determined by the
normal condion, we end the calculus with $K$-matrix solutions of (\ref{int.2}%
) with $N^{2}/2+1/2$ free parameters $\beta _{i,j}$ if $N$ is odd and with $%
N^{2}/2+1$ free parameters if $N$ is even.

Now, let us describe about the corresponding diagonal $K$-matrix solutions.

\subsection{The diagonal solutions}

Taking into account only the diagonal entries, the reflection equations are
solved when we find all matrix elements $k_{j,j}(u)$, $j=2,...,N$ as
function of $k_{1,1}(u)$, provided that the diagonal parameters $\beta
_{j,j} $ satisfy $(N-1)(N-2)/2$ constraint equations of the type%
\begin{equation}
\left( \beta _{N,N}-\beta _{i,i}\right) \left( \beta _{N,N}-\beta
_{j,j}\right) \left( \beta _{j,j}-\beta _{i,i}\right) =0\qquad (i\neq j\neq
N)  \label{red.1}
\end{equation}%
From (\ref{red.1}) we find solutions with only two type of entries. Let us
normalize one of them to be equal to $1$ such that the other one has the form%
\begin{equation}
k_{p,p}(u)=-\frac{\beta _{p,p}x_{2}(u)\left[ \Delta _{1}x_{2}(u)+x_{1}(u)%
\right] +2\left[ x_{1}(u)x_{2}^{\prime }(u)-x_{1}^{\prime }(u)x_{2}(u)\right]
}{\beta _{p,p}x_{2}(u)\left[ \Delta _{2}x_{2}(u)+x_{1}(u)\right] -2\left[
x_{1}(u)x_{2}^{\prime }(u)-x_{1}^{\prime }(u)x_{2}(u)\right] }  \label{red.2}
\end{equation}%
with $\Delta _{1}+\Delta _{2}=2\cosh \eta $.

Identifying the diagonal positions of the $K$-matrix with the matrix
elements of the $M$-matrix (\ref{M.1}--\ref{M.4}), $(1,2,...,N)\circeq
(M_{1,1},M_{2,2},...,M_{N,N})$ one can see that $\Delta _{1}$ is the sum of
the $M_{j,j}$ corresponding to the positions of the entries $1$ and $\Delta
_{2}$ is the sum of the $M_{j,j}$ corresponding to the positions of the
entries $k_{p,p}(u)$.

Denoting the diagonal solutions by ${\Bbb K}_{{\bf a}}^{[r]}$ where ${\bf a}%
=(a_{1},a_{2},...,a_{N})$ with $a_{i}=0$ if $k_{i,i}(u)=1$ or $a_{i}=1$ if $%
k_{i,i}(u)=k_{pp}(u)$ and $r$ is the number of the entries $k_{p,p}(u)$
distributed on diagonal positions and $p$ being the first position with the
entry different from $1$. Thus, we have counted

\begin{equation}
Z=\sum\limits_{r=1}^{N-1}\frac{N!}{r!\left( N-r\right) !}  \label{red.3}
\end{equation}%
for the number of diagonal $K^{-}$ matrix solutions with one free parameter.

The dual equation (\ref{int.3}) is solved by the $K^{+}$matrices via the
isomorphism (\ref{int.4}) with $\rho =-\eta $ and the matrix $M$ specified
by (\ref{M.1}--\ref{M.4}). Here we note that trace of each diagonal $M$%
-matrix is equal to $2\cosh \eta $.

Now, we explicitly show these computations for the first models. Before, we
can use the identity%
\begin{equation}
\frac{x_{2}(u)[x_{2}(u)\cosh \eta +x_{1}(u)]}{x_{1}(u)x_{2}^{\prime
}(x)-x_{1}^{\prime }(u)x_{2}(u)}=\sinh (u)\cosh (u)  \label{lkm.8}
\end{equation}%
in order to simplify our presentation. From the solution ((\ref{lkm.4}) to (%
\ref{k11})) one can see $k_{1,N}(u)$ as an arbitrary function satisfying the
normal condition. Therefore, the choice 
\begin{equation}
k_{1,N}(u)=\frac{1}{2}\beta _{1,N}\sinh (2u)  \label{k1N}
\end{equation}%
doesn't implies any restriction as compared to the general case.

\subsection{The $A_{1}^{(1)}$\ spin-$\frac{1}{2}$ and $C_{1}^{(1)}$
Temperley-Lieb K-matrices}

For these models we have the well-known three free parameter solution for
the ${\cal U}_{q}[sl(2)]$\ spin-$\frac{1}{2}$ model \cite{GZ, DeVega} 
\begin{equation}
K^{-}(u)=\left( 
\begin{array}{cc}
k_{1,1}(u) & \frac{1}{2}\beta _{1,2}\sinh (2u) \\ 
&  \\ 
\frac{1}{2}\beta _{2,1}\sinh (2u) & k_{1,1}(u)+\frac{1}{2}(\beta
_{2,2}-\beta _{1,1})\sinh (2u)%
\end{array}%
\right)  \label{lkm.9}
\end{equation}%
Using the identity (\ref{lkm.8}) and (\ref{k1N}), the expression for $%
k_{1,1}(u)$ (\ref{k11}) has a simplified form 
\begin{equation}
k_{1,1}(u)=1-\frac{1}{2}(\beta _{2,2}-\beta _{1,1})\left[ x_{1}(u)+q\
x_{2}(u)\right] x_{2}(u)\sinh \eta  \label{lkm.10}
\end{equation}%
where $\beta _{1,2}$, $\beta _{2,1}$ and $\beta _{2,2}$ being the free
parameters and $2\cosh \eta =q+q^{-1}$. Moreover, we find that the ${\cal U}%
_{q}[sp(2)]$ {\small TL} model has the same $K$-matrix form but, with%
\begin{equation}
k_{1,1}(u)=1-\frac{1}{2}(\beta _{2,2}-\beta _{1,1})\left[ x_{1}(u)+q^{2}\
x_{2}(u)\right] x_{2}(u)\sinh \eta
\end{equation}%
since that now $2\cosh \eta =q^{2}+q^{-2}$.

The entries of the diagonal solutions $k_{1,1}(u)$ and $k_{2,2}(u)$ are
given by (\ref{red.2}) and we have two solutions for each model%
\begin{eqnarray}
{\Bbb K}_{(1,0)}^{[1]} &=&\left( 
\begin{array}{cc}
k_{1,1}(u) & 0 \\ 
0 & 1%
\end{array}%
\right) ,\quad \Delta _{1}=M_{2,2}\quad \Delta _{2}=M_{1,1}\qquad  \nonumber
\\
{\Bbb K}_{(0,1)}^{[1]} &=&\left( 
\begin{array}{cc}
1 & 0 \\ 
0 & k_{2,2}(u)%
\end{array}%
\right) ,\quad \Delta _{1}=M_{1,1}\quad \Delta _{2}=M_{2,2}
\end{eqnarray}%
where $M_{1,1}=q^{-1}$, $M_{2,2}=q$ for ${\cal U}_{q}[sl(2)]$\ spin-$\frac{1%
}{2}$ model and $M_{1,1}=q^{-2}$, $M_{2,2}=q^{2}$ for ${\cal U}_{q}[sp(2)]$
model.\ Of course, in both models ${\Bbb K}_{(1,0)}^{[1]}$ and ${\Bbb K}%
_{(0,1)}^{[1]}$ are equivalent by the exchange $q\leftrightarrow q^{-1}$.

\subsection{The $A_{1}^{(1)}$ spin-$1$ Temperley-Lieb K matrices}

For the biquadratic model \cite{Batch4, KLS}, it follows from (\ref{lkm.4})
and (\ref{lkm.5}) that%
\begin{eqnarray}
K^{-}(u) &=&\left( 
\begin{array}{ccc}
k_{1,1}(u) & \frac{1}{2}\beta _{1,2}\sinh (2u) & \frac{1}{2}\beta
_{1,3}\sinh (2u) \\ 
&  &  \\ 
\frac{1}{2}\beta _{2,1}\sinh (2u) & k_{1,1}(u)+\frac{1}{2}(\beta
_{2,2}-\beta _{1,1})\sinh (2u) & \frac{1}{2}\beta _{2,3}\sinh (2u) \\ 
&  &  \\ 
\frac{1}{2}\beta _{3,1}\sinh (2u) & \frac{1}{2}\beta _{3,2}\sinh (2u) & 
k_{1,1}(u)+\frac{1}{2}(\beta _{3,3}-\beta _{1,1})\sinh (2u)%
\end{array}%
\right)  \nonumber \\
&&  \label{lkm.11}
\end{eqnarray}%
where $k_{1,1}(u)$ is given by (\ref{lkm.6}),%
\begin{eqnarray}
k_{1,1}(u) &=&1-\frac{1}{2}\left\{ (\beta _{3,3}-\beta _{1,1})\left[
x_{1}(u)+M_{3,3}x_{2}(u)\right] +x_{1}(u)\Psi _{1,3}\right.  \nonumber \\
&&\left. +(\beta _{2,2}-\beta _{1,1})M_{2,2}x_{2}(u)\right\} x_{2}(u)\sinh
\eta .  \label{lkm.12}
\end{eqnarray}%
The diagonal parameters are fixed by the constraint equations (\ref{diag.1})
and (\ref{diag.2})%
\begin{eqnarray}
\beta _{2,2} &=&\beta _{1,1}+\Psi _{1,3}-\Psi _{2,3}=\beta _{1,1}+\frac{%
\beta _{1,2}\beta _{2,3}}{\beta _{13}}-\frac{\beta _{21}\beta _{13}}{\beta
_{23}}\  \\
\beta _{3,3} &=&\beta _{1,1}+\Psi _{2,1}-\Psi _{2,3}=\beta _{1,1}+\frac{%
\beta _{1,3}\beta _{3,2}}{\beta _{1,2}}-\frac{\beta _{2,1}\beta _{1,3}}{%
\beta _{2,3}},\ 
\end{eqnarray}%
and $\beta _{11}$ is fixed by the normal condition. Moreover, all remained
constraint equations are solved by the relation%
\begin{equation}
\beta _{3,1}=\beta _{1,3}\frac{\beta _{3,2}\beta _{2,1}}{\beta _{1,2}\beta
_{2,3}}\qquad {\rm or}\qquad \Psi _{3,1}=\Psi _{1,3}
\end{equation}%
and we have get a five free parameter solution.

Here $2\cosh \eta =q^{-2}+1+q^{2}$. It means that $M_{11}=q^{-2}$, $M_{22}=1$
and $M_{33}=q^{2}$. Among several possibilities, we made the choice \ $\beta
_{1,2}$, $\beta _{1,3}$, $\beta _{2,1}$, $\beta _{2,3}$ and $\beta _{3,2}$\
\ for the free parameters.

The correspondig diagonal solutions are six, half of them with one entry
different from $1$%
\begin{equation}
{\Bbb K}_{(1,0,0)}^{[1]}=\left( 
\begin{array}{ccc}
k_{1,1}(u) & 0 & 0 \\ 
0 & 1 & 0 \\ 
0 & 0 & 1%
\end{array}%
\right) ,\quad \Delta _{1}=M_{2,2}+M_{3,3},\quad \Delta _{2}=M_{1,1}
\end{equation}%
\begin{equation}
{\Bbb K}_{(0,1,0)}^{[1]}=\left( 
\begin{array}{ccc}
1 & 0 & 0 \\ 
0 & k_{2,2}(u) & 0 \\ 
0 & 0 & 1%
\end{array}%
\right) ,\quad \Delta _{1}=M_{1,1}+M_{3,3},\quad \Delta _{2}=M_{2,2}
\end{equation}%
\begin{equation}
{\Bbb K}_{(0,0,1)}^{[1]}=\left( 
\begin{array}{ccc}
1 & 0 & 0 \\ 
0 & 1 & 0 \\ 
0 & 0 & k_{3,3}(u)%
\end{array}%
\right) ,\quad \Delta _{1}=M_{1,1}+M_{2,2},\quad \Delta _{2}=M_{3,3}
\end{equation}%
and three further diagonal solutions with two equal entries different from
unity%
\begin{equation}
{\Bbb K}_{(1,1,0)}^{[2]}=\left( 
\begin{array}{ccc}
k_{1,1}(u) & 0 & 0 \\ 
0 & k_{1,1}(u) & 0 \\ 
0 & 0 & 1%
\end{array}%
\right) \quad \Delta _{1}=M_{3,3},\quad \Delta _{2}=M_{1,1}+M_{2,2}
\end{equation}%
\begin{equation}
{\Bbb K}_{(1,0,1)}^{[2]}=\left( 
\begin{array}{ccc}
{\bf k}_{1,1}(u) & 0 & 0 \\ 
0 & 1 & 0 \\ 
0 & 0 & {\bf k}_{1,1}(u)%
\end{array}%
\right) \quad \Delta _{1}=M_{2,2},\quad \Delta _{2}=M_{1,1}+M_{3,3}
\end{equation}%
\begin{equation}
{\Bbb K}_{(0,1,1)}^{[2]}=\left( 
\begin{array}{ccc}
1 & 0 & 0 \\ 
0 & k_{2,2}(u) & 0 \\ 
0 & 0 & k_{2,2}(u)%
\end{array}%
\right) \quad \Delta _{1}=M_{1,1},\quad \Delta _{2}=M_{2,2}+M_{3,3}
\end{equation}%
where the entries $k_{p,p}(u)$ are given by (\ref{red.2}).

Here we notice again that the difference between the diagonal entries come
from the partitions of $\Delta _{1}+\Delta _{2}=q^{-2}+1+q^{2}$ and the
equivalence between them due to the symmetry $q\leftrightarrow q^{-1}$.

We also note that if $q^{2}$ is replaced by $q$, the equivalence $%
B_{1}^{(1)}\simeq A_{1}^{(1)}$ spin-$1$ is manifested in all expressions
above, since that $2\cosh \eta =q^{-1}+1+q$ for $B_{1}^{(1)}$.

These diagonal solutions were recently used in \cite{RLS2} to study the
spectrum of the spin-$1$ {\small TL} spin chain with integrable open
boundary conditions.

\subsection{The $A_{1}^{(1)}$ spin-$\frac{3}{2}$ and $C_{2}^{(1)}$
Temperley-Lieb K matrices}

For ${\cal U}_{q}[sl(2)]$ spin-$\frac{3}{2}$ and ${\cal U}_{q}[sp(4)]$
models, we have from (\ref{lkm.4}) to (\ref{lkm.6}) the following
non-diagonal entries%
\begin{equation}
k_{i,j}(u)=\frac{1}{2}\beta _{i,j}\sinh (2u),\quad (i\neq j=1,2,3,4)
\label{lkm.14}
\end{equation}%
and the diagonal one%
\begin{equation}
k_{i,i}(u)=k_{1,1}(u)+\frac{1}{2}(\beta _{i,i}-\beta _{1,1})\sinh
(2u),\qquad (i=2,3,4)  \label{lkm.15}
\end{equation}%
with 
\begin{eqnarray}
k_{1,1}(u) &=&1-\frac{1}{2}\left\{ (\beta _{4,4}-\beta _{1,1})\left[
x_{1}(u)+M_{4,4}x_{2}(u)\right] +x_{1}(u)\Psi _{1,4}\right.  \nonumber \\
&&\left. +[(\beta _{2,2}-\beta _{1,1})M_{2,2}+(\beta _{3,3}-\beta
_{1,1})M_{3,3}]x_{2}(u)\right\} x_{2}(u)\sinh \eta .  \label{lkm.16}
\end{eqnarray}%
where $M_{1,1}=q^{-3}$, $M_{2,2}=q^{-1}$, $M_{3,3}=q$ and $M_{4,4}=q^{3}$
for ${\cal U}_{q}[sl(2)]$ spin-$\frac{3}{2}$ model and $M_{1,1}=q^{-4}$, $%
M_{2,2}=q^{-2}$, $M_{3,3}=q^{2}$ and $M_{4,4}=q^{4}$ for ${\cal U}%
_{q}[sp(4)] $ model.

From the equations (\ref{diag.1}) and (\ref{diag.2}) we choose to fix the
following diagonal parameters:%
\begin{eqnarray}
\beta _{2,2} &=&\beta _{1,1}+\Psi _{1,4}-\Psi _{2,4}=\beta _{1,1}+\frac{%
\beta _{1,2}\beta _{2,4}+\beta _{1,3}\beta _{3,4}}{\beta _{1,4}}-\frac{\beta
_{2,1}\beta _{1,4}+\beta _{2,3}\beta _{3,4}}{\beta _{2,4}},  \nonumber \\
\beta _{3,3} &=&\beta _{1,1}+\Psi _{1,4}-\Psi _{3,4}=\beta _{1,1}+\frac{%
\beta _{1,2}\beta _{2,4}+\beta _{1,3}\beta _{3,4}}{\beta _{1,4}}-\frac{\beta
_{3,1}\beta _{1,4}+\beta _{3,2}\beta _{2,4}}{\beta _{3,4}},  \nonumber \\
\beta _{4,4} &=&\beta _{1,1}+\Psi _{1,3}-\Psi _{3,4}=\beta _{1,1}+\frac{%
\beta _{1,2}\beta _{2,3}+\beta _{1,4}\beta _{4,3}}{\beta _{1,3}}-\frac{\beta
_{3,1}\beta _{1,4}+\beta _{3,2}\beta _{2,4}}{\beta _{3,4}}.  \label{lkm.17}
\end{eqnarray}%
All remained constraint equations are solved by the choice%
\begin{eqnarray}
\beta _{4,1} &=&\frac{\beta _{4,2}\beta _{2,1}+\beta _{4,3}\beta _{3,1}}{%
\beta _{1,2}\beta _{2,4}+\beta _{1,3}\beta _{3,4}}\beta _{1,4}\qquad {\rm or}%
\qquad \Psi _{4,1}=\Psi _{1,4}  \nonumber \\
\beta _{3,2} &=&-\frac{\beta _{3,1}\beta _{1,2}+\beta _{3,4}\beta _{4,2}}{%
\beta _{1,2}\beta _{2,4}+\beta _{1,3}\beta _{3,4}}\beta _{1,4}\qquad {\rm or}%
\qquad \Psi _{3,2}=-\Psi _{1,4}  \nonumber \\
\beta _{2,3} &=&-\frac{\beta _{2,1}\beta _{1,3}+\beta _{2,4}\beta _{4,3}}{%
\beta _{1,2}\beta _{2,4}+\beta _{1,3}\beta _{3,4}}\beta _{1,4}\qquad {\rm or}%
\qquad \Psi _{2,3}=-\Psi _{1,4}  \label{lkm.18}
\end{eqnarray}%
It means that we have found a $K^{-}$ matrix with nine free parameters.

There are four diagonal solutions with one entry $k_{p,p}(u)$ and three
igual to unity%
\begin{eqnarray}
{\Bbb K}_{(1,0,0,0)}^{[1]} &:&\qquad \Delta _{1}=M_{1,1},\qquad \Delta
_{2}=M_{2,2}+M_{3,3}+M_{4,4}  \nonumber \\
&&\vdots  \nonumber \\
{\Bbb K}_{(0,0,0,1)}^{[1]} &:&\qquad \Delta _{1}=M_{4,4},\qquad \Delta
_{2}=M_{1,1}+M_{2,2}+M_{3,3}
\end{eqnarray}%
There are six diagonal solutions with two entry $k_{p,p}(u)$ and two igual
to unity%
\begin{eqnarray}
{\Bbb K}_{(1,1,0,0)}^{[2]} &:&\qquad \Delta _{1}=M_{1,1}+M_{2,2},\qquad
\Delta _{2}=M_{3,3}+M_{4,4}  \nonumber \\
&&\vdots  \nonumber \\
{\Bbb K}_{(0,0,1,1)}^{[2]} &:&\qquad \Delta _{1}=M_{3,3}+M_{4,4},\qquad
\Delta _{2}=M_{1,1}+M_{2,2}
\end{eqnarray}%
and more four solutions with three entries $k_{p,p}(u)$ and one equal to
unity%
\begin{eqnarray}
{\Bbb K}_{(1,1,1,0)}^{[3]} &:&\qquad \Delta
_{1}=M_{1,1}+M_{2,2}+M_{3,3},\qquad \Delta _{2}=M_{4,4}  \nonumber \\
&&\vdots  \nonumber \\
{\Bbb K}_{(0,1,1,1)}^{[3]} &:&\qquad \Delta
_{1}=M_{2,2}+M_{3,3}+M_{4,4},\qquad \Delta _{2}=M_{1,1}
\end{eqnarray}%
Remember that in these $14$ diagonal solutions we have $\Delta _{1}+\Delta
_{2}=q^{-3}+q^{-1}+q+q^{3}$ for the $A_{1}^{(1)}$ spin-$\frac{3}{2}$ model
and $\Delta _{1}+\Delta _{2}=q^{-4}+q^{-2}+q^{2}+q^{4}$ \ for the $%
C_{2}^{(1)}$ model.

\subsection{The $A_{1}^{(1)}$ spin-$2$ and $B_{2}^{(1)}$ Temperley-Lieb K
matrices}

\bigskip For $N=5$, the matrix elements are%
\begin{equation}
k_{i,j}(u)=\frac{1}{2}\beta _{i,j}\sinh (2u),\quad (i\neq j=1,...,5)
\end{equation}%
and%
\begin{equation}
k_{i,i}(u)=k_{1,1}(u)+\frac{1}{2}(\beta _{i,i}-\beta _{1,1})\sinh
(2u),\qquad (i=2,...,5).
\end{equation}%
where%
\begin{eqnarray}
k_{1,1}(u) &=&1-\frac{1}{2}\left\{ (\beta _{5,5}-\beta _{1,1})\left[
x_{1}(u)+M_{5,5}x_{2}(u)\right] +\Psi _{1,5}x_{1}(u)\right.  \nonumber \\
&&\left. +[(\beta _{2,2}-\beta _{1,1})M_{2,2}+(\beta _{3,3}-\beta
_{1,1})M_{3,3}+(\beta _{4,4}-\beta _{1,1})M_{4,4}]x_{2}(u)\right\}
x_{2}(u)\sinh \eta
\end{eqnarray}

The diagonal parameters are given by (\ref{diag.1}) 
\begin{eqnarray}
\beta _{2,2} &=&\beta _{1,1}+\Psi _{1,5}-\Psi _{2,5},\qquad \beta
_{3,3}=\beta _{1,1}+\Psi _{1,5}-\Psi _{3,5}  \nonumber \\
\beta _{4,4} &=&\beta _{1,1}+\Psi _{1,5}-\Psi _{4,5},\qquad
\end{eqnarray}%
and by (\ref{diag.2})%
\begin{equation}
\beta _{5,5}=\beta _{1,1}+\Psi _{1,4}-\Psi _{4,5}
\end{equation}

Moreover, we have $5(5-1)/2=10$ symmetric relations%
\begin{equation}
\Psi _{j,i}=\Psi _{i,j}\qquad (j>i)
\end{equation}%
more $2(5-3)=4$ relations com four $\Psi _{i,j}$ 
\begin{eqnarray}
\Psi _{2,4} &=&\Psi _{2,3}+\Psi _{1,4}-\Psi _{1,3},\qquad \Psi _{2,5}=\Psi
_{2,3}+\Psi _{1,5}-\Psi _{1,3},  \nonumber \\
\Psi _{3,4} &=&\Psi _{2,3}+\Psi _{1,4}-\Psi _{1,2},\qquad \Psi _{3,5}=\Psi
_{2,3}+\Psi _{1,5}-\Psi _{1,2},
\end{eqnarray}%
and $(5-3)(5-4)/2=1$ relation with six $\Psi _{i,j}$%
\begin{equation}
\Psi _{4,5}=\Psi _{1,5}+\Psi _{1,4}+\Psi _{2,3}-\Psi _{1,2}-\Psi _{1,3}.
\end{equation}

As mentioned above, theses $5(5-2)=15$ relations are enough to fix seven
remained parameters:%
\begin{eqnarray}
\beta _{2,1} &=&-\frac{\beta _{1,2}\beta _{2,3}\beta _{2,5}}{\beta
_{1,3}\beta _{1,5}}-\frac{\beta _{2,3}\beta _{3,5}}{\beta _{1,5}}-\frac{%
\beta _{2,5}\beta _{5,3}}{\beta _{1,3}}-\frac{\beta _{2,3}\beta _{4,5}(\beta
_{1,4}\beta _{2,5}-\beta _{1,5}\beta _{2,4})}{\beta _{1,5}(\beta _{1,3}\beta
_{2,5}-\beta _{1,5}\beta _{2,3})}  \nonumber \\
&&-\frac{\beta _{2,5}\beta _{4,3}(\beta _{1,3}\beta _{2,4}-\beta _{1,4}\beta
_{2,3})}{\beta _{1,3}(\beta _{1,3}\beta _{2,5}-\beta _{1,5}\beta _{2,3})}
\end{eqnarray}%
from which we can find $\beta _{3,1}$, $\beta _{4,1}$ and $\beta _{5,1}$,
replacing the indexes $(2\leftrightarrow 3)$, $(2\leftrightarrow 4)$ and $%
(2\leftrightarrow 5)$, respectively. From the parameter%
\begin{eqnarray}
\beta _{3,2} &=&\frac{\beta _{1,2}\beta _{1,4}\beta _{4,5}}{\beta
_{1,3}\beta _{1,5}}+\frac{\beta _{1,2}\beta _{3,5}}{\beta _{1,5}}-\frac{%
\beta _{1,4}\beta _{4,2}}{\beta _{1,3}}  \nonumber \\
&&+\frac{(\beta _{1,3}\beta _{2,5}-\beta _{1,5}\beta _{2,3})(\beta
_{1,2}\beta _{4,5}-\beta _{1,5}\beta _{4,2})}{(\beta _{1,3}\beta
_{4,5}-\beta _{1,5}\beta _{4,3})}\left\{ \frac{\beta _{1,2}}{\beta _{1,5}}%
\right.  \nonumber \\
&&+\left. \frac{\beta _{1,3}\beta _{5,4}-\beta _{1,4}\beta _{5,3}}{\beta
_{1,3}\beta _{2,4}-\beta _{1,4}\beta _{2,3}}\right\}
\end{eqnarray}%
we can also find $\beta _{5,2}$, replacing the indexes $(3\leftrightarrow 5)$%
. \ The last parameter is%
\begin{eqnarray}
\beta _{3,4} &=&\frac{\beta _{1,4}\beta _{3,5}}{\beta _{1,5}}+\frac{\beta
_{1,4}\beta _{2,5}-\beta _{1,5}\beta _{2,4}}{\beta _{1,3}\beta _{1,5}}%
\left\{ \frac{\beta _{1,4}(\beta _{1,3}\beta _{4,5}-\beta _{1,5}\beta _{4,3})%
}{\beta _{1,3}\beta _{2,5}-\beta _{1,5}\beta _{2,3}}\right.  \nonumber \\
&&+\left. \frac{\beta _{1,5}(\beta _{1,3}\beta _{5,4}-\beta _{1,4}\beta
_{5,3})}{\beta _{1,3}\beta _{2,4}-\beta _{1,4}\beta _{2,3}}+\beta
_{1,2}\right\}
\end{eqnarray}%
These seven parameters plus the five diagonal parameters $\beta _{i,i}$ give
us a $5$ by $5$ reflection matrix solution with $13$ free parameters.
Similarly, the corresponding $30$ diagonal solutions can be written for both
models.

In the sequence ($N\geq 6$), the expressions of the fixed parameters are too
large and cumbersome.

\subsection{A reduced solution}

An important characteristic of the {\small TL} boundary solutions is its
large number of free parameters. It means that we have many different
reduced solutions for a given ${\cal R}$-matrix. In particular, choosing the
free parameters one can gets an appropriate $K$-matrix solution. For
instance, if we consider all $\beta _{i,j}=\beta \ (i\neq j)$ and all $\beta
_{i,i}=\alpha $, in (\ref{k11}), we will get one-parameter solution of the
form%
\begin{equation}
K(u)=f_{1}(u)I+f_{2}(u)G  \label{baxt}
\end{equation}%
where 
\begin{eqnarray}
f_{1}(u) &=&1-\frac{N-2}{2}\beta x_{1}(u)x_{2}(u)\sinh (\eta )  \nonumber \\
f_{2}(u) &=&\frac{1}{2}\beta \sinh (2u).
\end{eqnarray}%
and \ $G$ is a $N$ by $N$ matrix with entries 
\begin{equation}
G_{i,j}=\left\{ 
\begin{array}{c}
0\qquad {\rm if}\qquad i=j \\ 
1\qquad {\rm if}\qquad i\neq j%
\end{array}%
\right.
\end{equation}%
Although $G$ satifies a quadratic relation as (\ref{int.5}) 
\begin{equation}
G^{2}-(N-2)G-(N-1)I=0,\qquad
\end{equation}%
we don't know as the solution (\ref{baxt}) can be fitted with (\ref{int.6}).
But, certainly, the infinity spectral parameter limit of the solutions
presented above will solve the constant reflection equations \cite{KSkly}.

Many other reduced solutions can be derived in a similar way. See, for
instance, the cases presented in \cite{LS3} for the ${\cal U}_{q}[sl(2)]$\
model.

\section{Conclusion}

In this work we have presented solutions of the reflection equation for the 
{\small TL} vertex models. Our findings can be summarized into two classes
of solutions depending on $N$-parity. A large number of free parameters is
an important caracteristic of these solutions.

In analogy com the $R$-matrices form (\ref{mod.1}), the $K$-matrices have
the same form (\ref{lkm.4}) and (\ref{lkm.5}), with all quantum group
dependence in the diagonal entries through of matrix elements of $M$.

These results pave the way to construct, solve and study physical properties
of the underlying quantum spin chains with open boundaries, generalizing the
previous efforts made for the case of periodic boundary conditions \cite%
{KLS2, LS2}.

We expect that the coordinate Bethe ansatz for all diagonal solutions
presented here can be obtained by adapting the results of \cite{RLS2, LSG}
and that its generalization, as in \cite{Crampe}, may be a possibility to
treat the non-diagonal solutions. We also expect that these results give
rise for the further developments on the subject of integrable open
boundaries for the {\small TL}vertex models based on superalgebras \cite%
{Zhang}.

\vspace{.55cm}%

{\bf Acknowledgment:} This work was supported in part by Funda\c{c}\~{a}o de
Amparo \`{a} Pesquisa do Estado de S\~{a}o Paulo ({\small FAPESP)}, grant
\#2011/18729-1 and by Conselho Nacional de Desenvolvimento-{\small CNPq}%
-Brasil.

\newpage%

\end{document}